\documentclass{mem}

\usepackage{natbib}
\usepackage{txfonts}
\usepackage{balance}
\usepackage{graphicx}
\usepackage[a4paper,breaklinks,dvipdfm]{hyperref}
\usepackage{amssymb}
\usepackage{fixltx2e}

\newcommand{\sevn}{{\sc SEVN}}
\newcommand{\parsec}{\textsc{PARSEC}}
\newcommand{\mzams}{\ensuremath{M_{\mathrm{ZAMS}}}}
\newcommand{\msun}{\ensuremath{{\rm M}_{\sun{}}}}
\newcommand{\zsun}{\ensuremath{{\rm Z}_{\sun{}}}}
\newcommand{\mbhmax}{\ensuremath{M_{\mathrm{BH,max}}}}
\newcommand{\mco}{\ensuremath{M_{\mathrm{CO}}}}

\begin{document}
\def\teff{$T\rm_{eff }$}
\def\kms{$\mathrm {km s}^{-1}$}
	
\title{Shedding light on the black hole mass spectrum}	

\subtitle{}
	
\author{M. \,Spera\inst{1} \and N. \,Giacobbo\inst{2} \and M. \,Mapelli\inst{1,3}}
	
\institute{ Istituto Nazionale di Astrofisica --
		Osservatorio Astronomico di Padova, Vicolo dell'Osservatorio 5,
		I-35122 Padova, Italy, \email{mario.spera@oapd.inaf.it}
		\and
		Universita degli Studi di Padova, Vicolo dell’Osservatorio 3, 
		I-35122, Padova, Italy
		\and 
		INFN, Milano Bicocca, Piazza della Scienza 3, I-20126, Milano, Italy
		}
	
\authorrunning{Spera, Giacobbo \& Mapelli}
\titlerunning{Black hole mass spectrum}

\abstract{The mass spectrum of stellar black holes (BHs) is highly uncertain. 
	Theoretical models of BH formation strongly depend on the efficiency of stellar winds of the progenitor star and on the supernova (SN) explosion mechanism. We discuss the BH mass spectrum we obtain using \sevn{}, a new public population-synthesis code that includes up-to-date stellar-wind prescriptions  and several SN explosion models. Our models indicate a sub-solar metallicity environment for the progenitors of the gravitational wave source GW150914. We show that our models predict substantially larger BH masses (up to $\sim 100\,{}\msun{}$) than other population synthesis codes, at low metallicity. In this proceeding, we also discuss the impact of pair-instability SNe on our previously published models.
		
}
	
\maketitle{}
	
\section{Introduction}
Stellar-mass black holes (BHs) 
play a key role on  a plethora of astrophysical processes, such as gravitational wave (GW) emission and X-ray binaries. Despite their crucial importance in astrophysics, the mass spectrum of BHs is still matter of debate. From the observational point of view the confirmed stellar-mass BHs are few tens and accurate dynamical mass measurements are available only for $\sim 20$ objects. Dynamically measured BH masses in the  Milky Way are all smaller than $\sim 15\,{}\msun{}$ \citep{ozel2010}. Still, the recent detection of the gravitational wave (GW) signal emitted by two merging BHs with masses $29^{+4}_{-4}$ M$_{\sun{}}$ and $36^{+5}_{-4}$ M$_{\sun{}}$ \citep{abbott2016a, abbott2016b} proves that massive BHs (i.e. BHs with mass $>25$ M$_{\sun{}}$, \citealt{mapelli2009}) exist. 

\begin{figure*}[t!]
	\resizebox{\hsize}{!}{\includegraphics[clip=true]{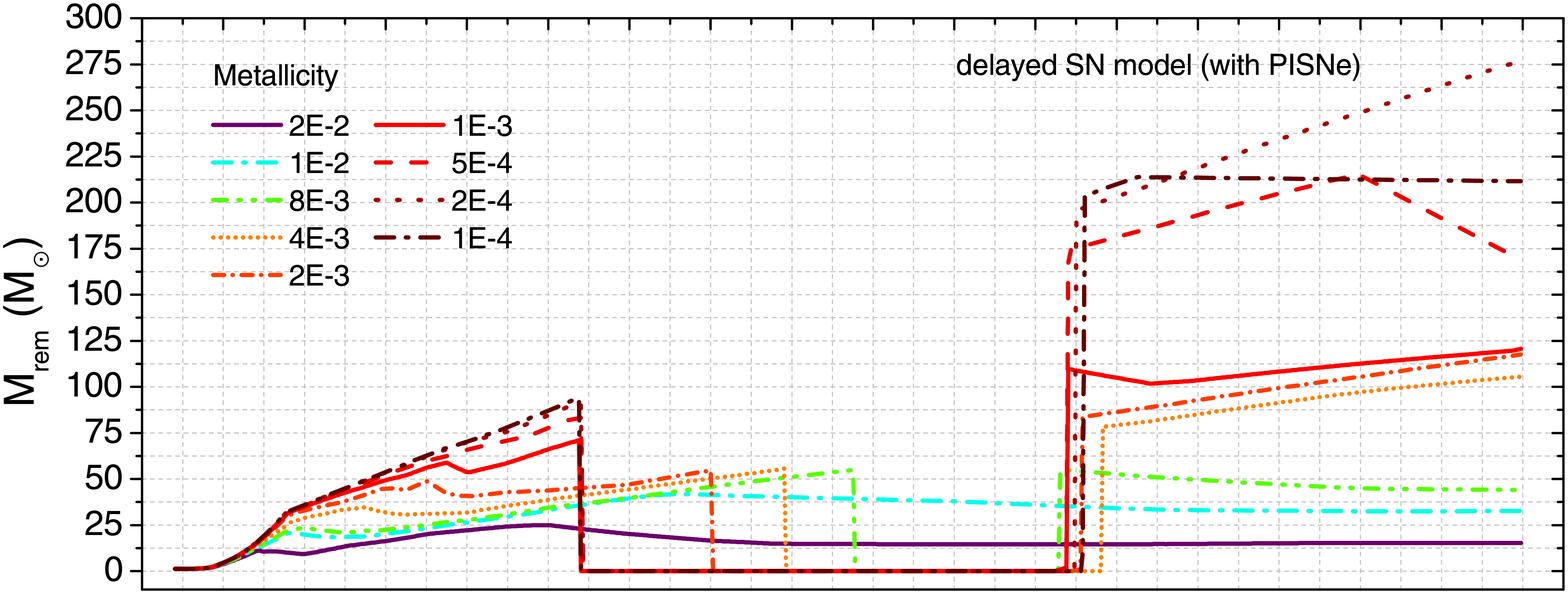}}
	\resizebox{\hsize}{!}{\includegraphics[clip=true]{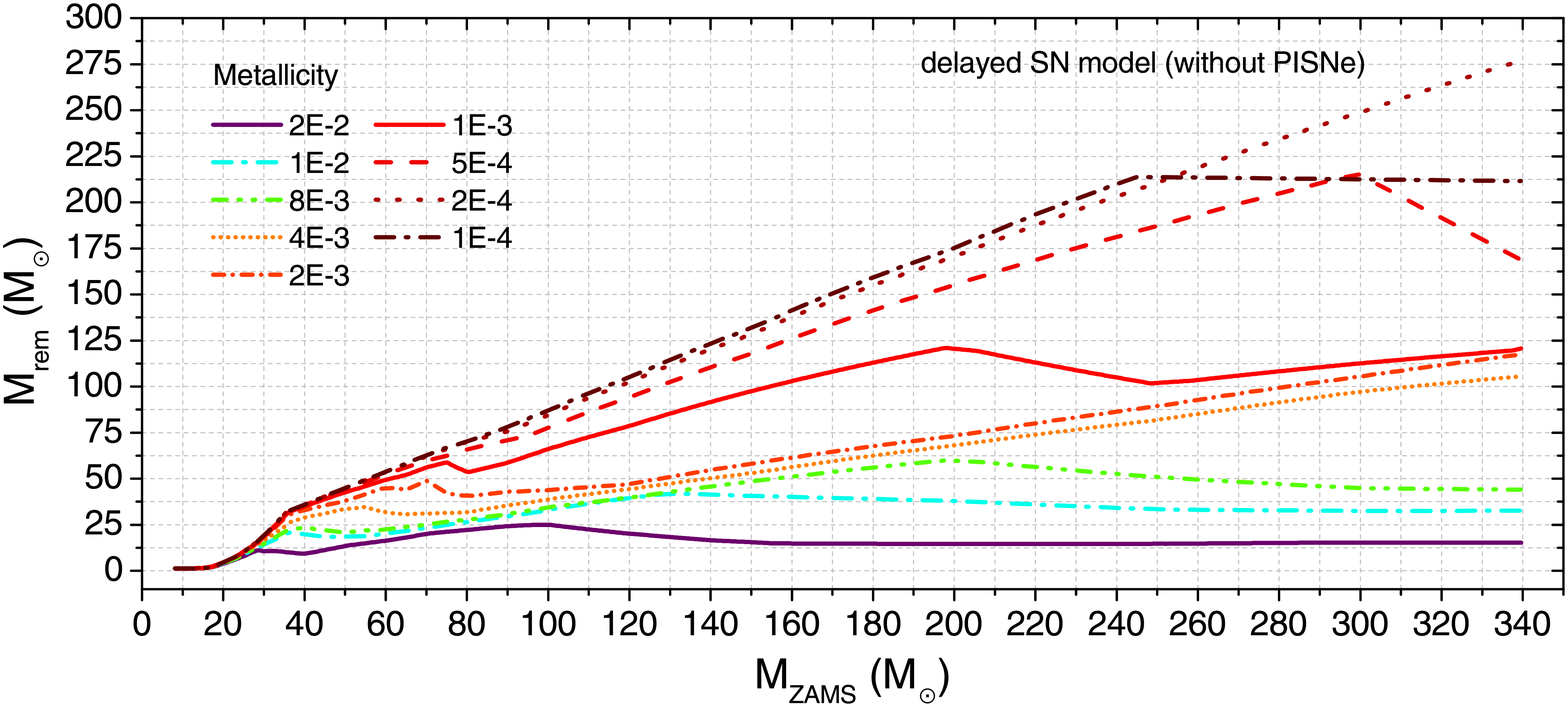}}
	\caption{\footnotesize Mass of the compact remnant ($M_{\mathrm{rem}}$) as a function of the initial mass of the star ($\mzams{}$), for various metallicities. The curves have been obtained using \sevn{} and the delayed SN explosion model. Top (bottom) panel: PISNe are (not) included. } 
	\label{fig1}
\end{figure*}

From a theoretical point of view, the link between progenitor stars and their compact remnants depends on two fundamental ingredients: (i) the mass loss of massive stars through stellar winds; (ii) the supernova (SN) explosion mechanism. Both the models of SN explosion (e.g. \citealt{fryer2012,ertl2016}) and the theory of massive star evolution (e.g. \citealt{tang2014}) were deeply revised in the last few years, especially for what concerns the evolution of massive stars ($m>30$ M$_{\sun{}}$) at low metallicity ($Z<0.1$ Z$_{\sun{}}$). For these reasons, population synthesis codes that aim at studying the BH mass spectrum must account for up-to-date models for both SN explosions and stellar evolution. 

Here we discuss the BH mass spectrum we obtained with \sevn{} (acronym for `Stellar EVolution for N-body codes'), a population-synthesis tool  we recently developed \citep{spera2015}. 
In particular, we describe a new version of \sevn{}, which includes a simple treatment of pair-instability supernovae (PISNe). 

\section{Description of \sevn{}}
\sevn{} is a new tool that couples up-to-date stellar evolution recipes with several SN explosion prescriptions to study the link between progenitor stars and their compact remnants. \sevn{} includes stellar evolution recipes by means of tabulated stellar isochrones, for a grid of masses and metallicities. \sevn{} reads and interpolates the input tables on the fly. This strategy makes \sevn{} versatile. By default, \sevn{} includes the \parsec{} stellar evolution isochrones \citep{bressan2012,tang2014,chen2014}.

\sevn{} implements five SN explosion models. Three of them are described in detail by \citet{fryer2012}: (i) the model implemented in the {\sc StarTrack} population-synthesis code (see \citealt{belc2008}); (ii) the {\em rapid} supernova model; (iii) the {\em delayed} supernova model. In these models, the mass of the compact remnant depends only on the final properties of the progenitor star, by means of the final Carbon-Oxygen core mass (\mco{}) and of the final (pre-SN) mass of the star. We also included two more sophisticated SN explosion recipes, based on the compactness of the star at the edge of the iron core  (described in \citealt{oconnor2011} and \citealt{ertl2016}). 

\sevn{} can be coupled with several N-body codes, including {\sc starlab} \citep{portegies2001} and {\sc higpus} \citep{capuzzo2013}. More details about \sevn{} can be found in \cite{spera2015}. 


In this proceeding we describe a revised version of \sevn{}, where we implemented recipes for pair-instability supernovae (PISNe). Specifically, we assume that stars with $\mco{} \gtrsim 45\msun{}$ (Helium mass $\gtrsim 65\msun{}$) undergo a PISN and disintegrate, without leaving any compact remnant. Stars with Helium mass above $\sim 135\msun{}$ collapse directly into a BH \citep{heger2003}. 


\section{Results}
Fig. \ref{fig1} shows the mass spectrum of compact remnants as a function of the zero-age main sequence mass \mzams{} of their progenitors (up to 350 M$_\odot$), for different values of metallicity. To obtain the curves shown in Fig. \ref{fig1}, we used the delayed SN model \citep{fryer2012}. In the top panel we also include the effect of PISNe \citep{heger2003}, which is omitted in the bottom panel. 
Thus, Fig. \ref{fig1} is an updated version of Fig. 6 of \citet{spera2015}, in which PISNe were not included, and only stars with  \mzams{} up to $150\msun{}$ were considered.


Fig. \ref{fig1} confirms that the lower the metallicity is, the higher the mass of the heaviest compact remnant. The maximum BH mass (\mbhmax{}) is $\sim 25\msun{}$ and $\sim 275\msun{}$,  at $Z=\zsun{}$ and $Z= 0.01\zsun{}$ (where $\zsun{}=0.01524$), respectively. 
PISNe occurr at $Z\leq{} 0.008\simeq 0.5\zsun{}$, in the range $175\msun{}\leq \mzams{} \leq 230\msun{}$. We notice that while the \mzams{} upper limit of the PISN window does not depend significantly on metallicity (its value is always $\sim 230\msun{}$), the lower limit goes from $\sim 175\msun{}$ at $Z\simeq 0.008$ to $\sim 110\msun{}$ for $Z\lesssim 0.001$. 

If PISNe are included, the heaviest BHs that form in a stellar population 
with $m_{\mathrm{ZAMS}}\le{}150\msun{}$ is $\sim 100 \msun{}$ at $Z\lesssim 2.0\times 10^{-4}$. If PISNe are not accounted for, the maximum BH mass for $m_{\mathrm{ZAMS}}\le{}150\msun{}$ is $\sim 130\msun{}$ at $Z\lesssim 2.0\times 10^{-4}$ \citep{spera2015}.

\begin{figure}[t!]
	\resizebox{\hsize}{!}{\includegraphics[clip=true]{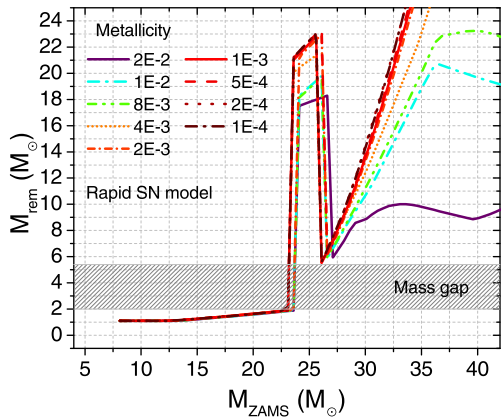}}
	\caption{\footnotesize Mass of the final compact remnant ($M_{\mathrm{rem}}$) as a function of the initial mass of the star ($\mzams{}$), for various metallicities, in the range $\mzams \lesssim 40\msun{}$. The curves have been obtained using \sevn{} and the rapid SN explosion model. The mass gap between the heaviest neutron star and the lightest BH (from $\sim 2 \msun{}$ to $\sim 5 \msun{}$) is highlighted by a shaded area. Line types are the same as in Fig. \ref{fig1}.}
	\label{fig2}
\end{figure}

Fig. \ref{fig2} shows the BH mass spectrum obtained with the rapid SN mechanism in the range $\mzams \lesssim 40\msun{}$. For $\mzams{} \gtrsim 40\msun{}$, the BH mass spectrum we obtain from the rapid SN model is the same as that obtained using the delayed SN explosion  (in this range all the stars collapse directly into a BH). The rapid and delayed SN models differ only at $\mzams{} < 40\msun{}$. The main feature of the rapid SN model is the abrupt step in the range $23\msun{}\lesssim \mzams{}\lesssim 28\msun{}$. In this window (which corresponds to $6\msun{}\leq \mco{} \leq 7\msun{}$) the rapid SN model predicts direct collapse into a BH \citep{fryer2012}. This implies that using the rapid SN model we do not form remnants with masses between $\sim 2\msun{}$ and $\sim 5\msun{}$. This agrees with current observations, which suggest a gap between the heaviest neutron star (NS) and the lightest BH \citep{ozel2010}, even though it is still unclear if this gap is physical or due to observational biases.

\begin{figure}[t!]
	\resizebox{\hsize}{!}{\includegraphics[clip=true]{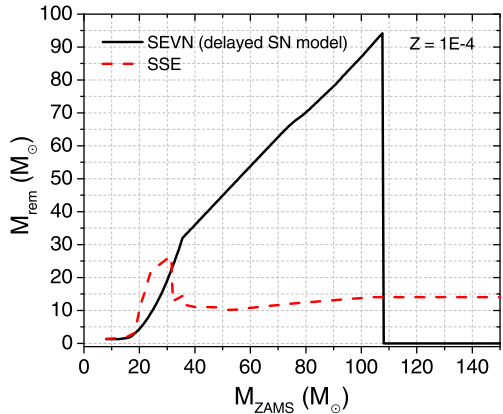}}
	\caption{\footnotesize Mass of compact remnants as a function of \mzams{} at $Z=10^{-4}$ derived with \sevn{} (solid black line) and  SSE (red dashed line). For SEVN, we adopted the delayed SN explosion mechanism.}
	\label{fig3}
\end{figure}
Fig. \ref{fig3} shows a comparison between the BH mass spectrum obtained with \sevn{} and that obtained using the single stellar evolution (SSE) population-synthesis code \citep{hurley2000}. The curves of Fig. \ref{fig3} are for $Z=10^{-4}$ and for \sevn{} we used the delayed SN explosion model. 
The differences in the BH mass spectrum between \sevn{} and SSE reflect the different stellar-wind prescriptions. 
\section{Summary}
The mass spectrum of BHs is still matter of debate. Observational constraints are few and theoretical models are affected by the uncertainties on SN explosion models and on the evolution of massive stars, especially at low metallicity. In this work we present the BH mass spectrum we obtained using \sevn{} \citep{spera2015}, a new population synthesis tool that couples up-to-date stellar evolution recipes (taken from the \parsec{} stellar evolution code \citealt{bressan2012,chen2014,tang2014}) and up-to-date SN explosion models (\citealt{oconnor2011,fryer2012,ertl2016}). PISNe are also accounted for in the current version of \sevn{}. \sevn{} predicts substantially larger BH masses at low metallicity than previous population synthesis codes. For a maximum stellar mass 
$\mzams{}=150\,{}\msun{}$, the maximum BH mass is $\sim 25$, $\sim 60$ and $\sim 100$ \msun{} at $Z=0.02$, $0.002$ and $0.0002$, respectively (when PISNe are accounted for). If we consider stars with \mzams{} up to $350\msun{}$ we can form BH with masses up to $\sim 275\msun{}$. 
These predicted BH masses have important implications for GW detections. In a forthcoming study, we will use \sevn{} to investigate the demographics of BH-BH binaries in star clusters.
\begin{acknowledgements}
We thank Alessandro Bressan for useful discussions. MS and MM acknowledge financial support from MIUR through grant FIRB 2012 RBFR12PM1F, from INAF through grant PRIN-2014-14, and  from the MERAC Foundation.
\end{acknowledgements}
\bibliographystyle{aa}

\end{document}